\documentclass[twocolumn,showpacs,amsmath]{revtex4}%
\usepackage{amsfonts}
\usepackage{amsmath}
\usepackage{amssymb}
\usepackage{graphicx}
\usepackage{appendix}
\usepackage{bibmods}
\usepackage{color}%
\providecommand{\U}[1]{\protect\rule{.1in}{.1in}}

\begin{document}
\title{Effects of thermal motion on electromagnetically induced absorption}
\author{E. Tilchin}
\affiliation{Department of Chemistry, Bar-Ilan University, Ramat Gan 52900, Israel}
\author{O. Firstenberg}
\affiliation{Department of Physics, Technion-Israel Institute of Technology, Haifa 32000, Israel}
\author{A. D. Wilson-Gordon}
\affiliation{Department of Chemistry, Bar-Ilan University, Ramat Gan 52900, Israel}


\begin{abstract}
We describe the effect of thermal motion and buffer-gas collisions on a
four-level closed $N$ system interacting with strong pump(s) and a weak probe.
This is the simplest system that experiences electromagnetically induced
absorption (EIA) due to transfer of coherence via spontaneous emission from
the excited to ground state. We investigate the influence of Doppler
broadening, velocity-changing collisions (VCC), and phase-changing collisions
(PCC) with a buffer gas on the EIA spectrum of optically active atoms. In
addition to exact expressions, we present an approximate solution for the
probe absorption spectrum, which provides physical insight into the behavior
of the EIA peak due to VCC, PCC, and wave-vector difference between the pump
and probe beams. VCC are shown to produce a wide pedestal at the base of the
EIA peak, which is scarcely affected by the pump-probe angular deviation,
whereas the sharp central EIA peak becomes weaker and broader due to the
residual Doppler-Dicke effect. Using diffusion-like equations for the atomic
coherences and populations, we construct a spatial-frequency filter for a
spatially structured probe beam and show that Ramsey narrowing of the EIA peak
is obtained for beams of finite width.

\end{abstract}
\maketitle

\section{\label{sec:intro}Introduction}

The absorption spectrum of a weak probe, interacting with a pumped
nearly-degenerate two-level transition, can exhibit either a sharp subnatural
dip or peak at line center \cite{Khitrova1988JOSAB}, depending on the
degeneracy of the levels, the polarizations of the fields, and the absence or
presence of a weak magnetic field. The phenomenon is termed
electromagnetically-induced transparency (EIT)
\cite{HarrisToday,Fleischhauer2005RMP} when there is a dip in the probe
spectrum and electromagnetically induced absorption (EIA)
\cite{Akulshin1998PRA} when there is a peak.

\begin{figure}
[ptb]
\begin{center}
\includegraphics[
height=2.2399in,
width=2.8078in
]%
{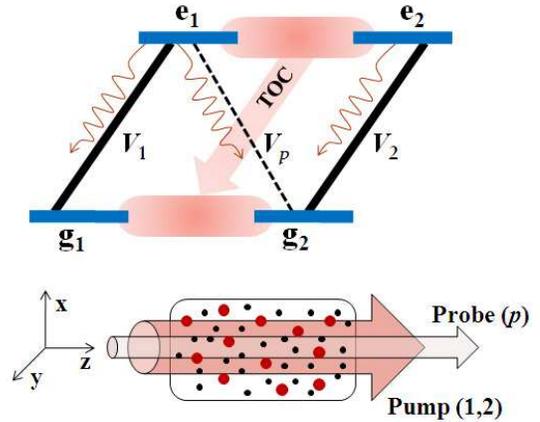}%
\caption{Top: The $N-$configuration atom. The light-induced transitions are
marked by solid (pump field) and dashed (probe field) lines, and the wavy
arrows are spontaneous decay paths. The thick arrow illustrates the
spontaneous transfer of coherence (TOC). Bottom: probe and pump beam(s), of
possibly a finite size, propagating through the vapor cell. The optical axis
is parallel to $\hat{z}$, while $\hat{x}$ and $\hat{y}$ form the optical
transverse plane.}%
\label{Fig. 1}%
\end{center}
\end{figure}

In the case of orthogonal polarizations of the pump and probe, both EIT and
EIA are related to the ground-level Zeeman coherence, which is induced by the
simultaneous action of both fields. The simplest model system that exhibits
EIT is the three-level $\Lambda$ system, where the two lower states $g_{1,2}$
are Zeeman sublevels of the ground hyperfine level $F_{g}$. In a $\Lambda$
system, quantum coherence can lead to the destructive interference between the
two possible paths of excitation. As a result, if the pump field is tuned to
resonance, the narrow dip in the probe absorption spectrum at the two-photon
resonance can be interpreted as EIT caused by a coherent population trapping
\cite{ArimondoCPTRev2} in the lower levels. The simplest system that exhibits
EIA is the four-level $N$ system \cite{Taichenachev1999PRA,GorenN2004PRA}
(Fig. \ref{Fig. 1}, top), consisting of states $g_{1,2}$ and $e_{1,2}$ which
are Zeeman sublevels of the ground ($F_{g}$) and excited ($F_{e}$) hyperfine
levels, where the $g_{i}\leftrightarrow e_{i}$, $i=1,2$, transitions interact
with non-saturating pump(s), and the $g_{2}\leftrightarrow e_{1}$ transition
interacts with a weak probe. The $N$ system gives similar results to those
obtained for a closed alkali-metal $F_{g}\rightarrow F_{e}=F_{g}+1$ transition
interacting with a $\sigma_{\pm}$ polarized pump, and a weak $\pi$ polarized
probe \cite{GorenN2004PRA,Zigdon2007SPIE}. It has been shown
\cite{Taichenachev1999PRA,GorenTOP2003PRA,GorenN2004PRA}, that the EIA peak is
due to transfer of coherence (TOC) from the excited state to the ground state,
via spontaneous emission. The excited-state coherence only exists in systems
where the coherent population trapping is incomplete so that there is some
population in the excited state \cite{GorenTOP2003PRA,Meshulam2007OL}. The
transfer of this coherence to the ground state leads to a peak in the
contribution of the ground-state two-photon coherence to the probe absorption
at line center, instead of the dip that occurs in its absence (for example, in
a $\Lambda$ system or a non-degenerate $N$ system) \cite{Zigdon2008PRA}.%

In this paper, we investigate the effect of the thermal motion of the
alkali-metal gas on the EIA spectrum, in the presence of a buffer gas. In a
previous paper \cite{GorenTOP2004PRA}, we discussed the effect of
phase-changing collisions (PCC) with the buffer gas on an $N$ system and
showed that they lead to considerable narrowing of the EIA peak in both the
presence and absence of Doppler broadening. These collisions increase the
transverse decay rate of the optical transitions, resulting in the so-called
pressure broadening of the optical spectral line, and are thus easily
incorporated in the Bloch equations. However, in order to describe the overall
effect of buffer-gas collisions, it is necessary to include both
velocity-changing collisions (VCC) as well as PCC
\cite{Singh1988JOSAB,May1999PRA}, which is a much greater challenge. Due to
the complexity of the problem, we limit our discussion to a four-level $N$
system, and to buffer-gas pressures that are sufficiently low so that
collisional decoherence of the excited state \cite{Failache2003PRA} can be neglected.

The Doppler effect occurs in the limit of \emph{ballistic} atomic motion, when
the mean free-path between VCC is much larger than the radiation wavelength.
Due to their narrow spectral response, Raman processes such as EIT and EIA are
much more sensitive to the ``residual" Doppler effect, arising when there is a
difference between the wavevectors of the Raman fields. In many cases however,
the Raman wavelength can become much larger than the typical free-path between
collisions. For example, an angular deviation of a milliradian between the two
optical beams yields a superposition pattern with a wavelength in the order of
a millimeter. In this limit, the atoms effectively perform a \emph{diffusion}
motion through the spatial oscillations of the superposition field, leading to
the Dicke narrowing of the residual Doppler width. While the residual Doppler
broadening is linearly proportional to the Raman wavevector, Dicke narrowing
shows a quadratic dependence. This behavior was demonstrated in EIT with
non-collimated pump and probe \cite{Weitz2005PRA,Shuker2007PRA}.

Recently, a model describing thermal motion and collisions for EIT was
presented \cite{Firstenberg2007PRA,Shuker2007PRA,Firstenberg2008PRA},
utilizing the density matrix distribution in space and velocity with a
Boltzmann relaxation formalism. The model describes a range of motional
phenomena, including Dicke narrowing, and diffusion in the presence of
electromagnetic fields and during storage of light. This diffusion model was
used to describe a spatial frequency filter for a spatially structured probe
\cite{Firstenberg2008PRA} and also Ramsey narrowing
\cite{Xiao2006PRL,Xiao2008OE}. Here, we utilize a similar formalism to
estimate the influence of the atomic thermal motion in a buffer-gas
environment, including VCC and PCC, on the spectral shape of EIA in a
four-level \textit{N} system, with collimated or non-collimated light beams.
In Sec. \ref{sec:DD}, the Doppler broadening and Dicke narrowing effects are
studied for plane-wave fields. As the full mathematical treatment is lengthy,
it appears in Appendix A. However, an approximate equation which describes the
main features of the spectra is presented in Sec. \ref{sec:DD}. Diffusion-like
equations for the ground and excited state coherences and populations are
derived in Appendix B. Two main phenomena are described using this model: (i)
a spatial-frequency filter for structured probe fields which is presented in
Sec. \ref{sec:filter}, and (ii) atomic diffusion through a finite-sized beam
resulting in Ramsey narrowing of the EIA peak, which is discussed in Sec.
\ref{sec:Ramseynarr}. Finally, conclusions are drawn in Sec. \ref{sec:conc}.

\section{\label{sec:DD}The Doppler-Dicke line shapes of EIA}

Consider the near-resonant interaction of a four-state atom in an \textit{N}
configuration, depicted in Fig. \ref{Fig. 1}. The two lower states $g_{1}$ and
$g_{2}$ are degenerate and belong to the ground level with zero energy, and
the excited states $e_{1}$ and $e_{2}$ are degenerate with energy $\hbar
\omega_{0}$. The light field consists of three beams, each with a carrier
frequency $\omega_{j}$ and wavevector $\mathbf{q}_{j},$ where $j=1,2$ denotes
the two strong pump beams, and $j=p$ the weak probe,%

\begin{equation}
\mathbf{\breve{E}}(\mathbf{r},t)=\sum_{j=1,2,p}\mathbf{E}_{j}\left(
\mathbf{r},t\right)  e^{-i\omega_{j}t+i\mathbf{q}_{j}\cdot\mathbf{r}%
}+\text{c.c.} \label{Eq. 1}%
\end{equation}
Here, $\mathbf{E}_{j}\left(  \mathbf{r},t\right)  $ are the slowly varying
envelopes in space and time. The pumps drive the $g_{1}\leftrightarrow e_{1}$
and $g_{2}\leftrightarrow e_{2}$ transitions, and the probe is coupled to the
$g_{2}\leftrightarrow e_{1}$ transition.%
\begin{figure}
[ptb]
\begin{center}
\includegraphics[
height=3.2553in,
width=3.4686in
]%
{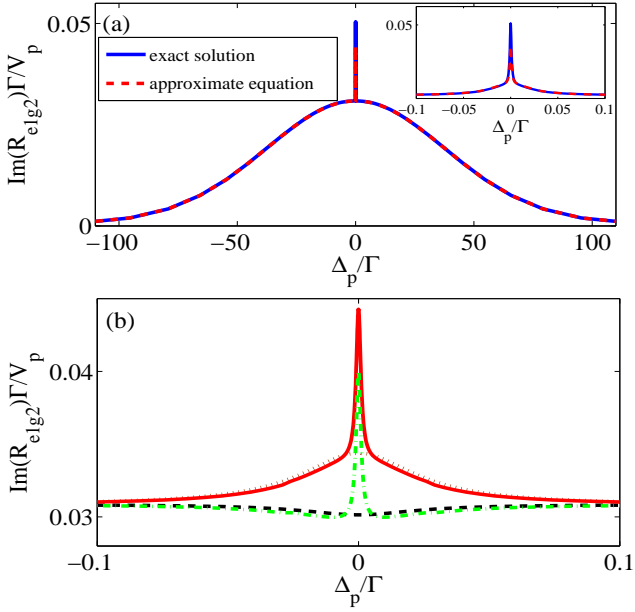}%
\caption{(a) The probe absorption, calculated from the exact solution for the
density matrix (blue line) and from the approximate solution Eq. (\ref{Eq. 2})
(red line), for collinear plane-wave beams ($\mathbf{q}_{1}=\mathbf{q}%
_{2}=\mathbf{q}_{p}$), $\Gamma_{\text{pcc}}=5\Gamma,$ $\gamma_{\text{vcc}%
}=0.025\Gamma$ $,$ $A=0.816,$ $V_{2}=0.1\Gamma,$ $V_{1}=AV_{2},$ $V_{p}%
=\gamma=0.001\Gamma,$ and $\Delta_{1}=\Delta_{2}=0$. The inset depicts a zoom
on the EIA peak with its wide pedestal. (b) The EIA\ spectrum (red line) is
the sum of three contributions in Eq. (\ref{Eq. 2}): the one-photon absorption
(black dashed line), a pedestal at the base of the peak (brown dotted line),
and the sharp peak (green dash-dotted line). For the clarity of presentation,
the one-photon absorption is added to the pedestal and to the sharp peak
curves.}%
\label{Fig. 2}%
\end{center}
\end{figure}

Our model will incorporate four relaxation rates: $\Gamma,$ the spontaneous
emission rate from the each of the excited states to all the ground states;
$\Gamma_{\text{pcc}},$ the pressure broadening of the optical transitions
resulting from PCC; $\gamma_{\text{vcc}}$, the velocity autocorrelation
relaxation rate ($1/\gamma_{\text{vcc}}$ is the time it takes the velocity
vector to vary substantially) \cite{Sobelman1967SPU}, which is proportional to
the rate of VCC; and $\gamma$ is the homogenous decoherence rate within the
ground and excited state manifolds due, for example, to spin-exchange and
spin-destruction collisions \footnote{The fact that the `inner' decoherence
rate $\gamma$ is shared by both the ground and the excited manifolds, does not
imply that their total decoherence rate is the same; the coherence between the
two excited states decays via the $e\rightarrow g$ relaxation channels and
therefore decays much faster than the ground-state coherence. Incorporating
different values of $\gamma$ for the ground and excited states does not lead
to substantial changes in the collision-induced phenomena explored here.}. In
the model, the transition $g_{1}\leftrightarrow e_{2}$ is forbidden (due to
some selection rule such as angular momentum).

To focus the discussion, we assume that all three beams are continuous waves,
namely $\mathbf{E}_{j}(\mathbf{r},t)=\mathbf{E}_{j}(\mathbf{r})$. We then
obtain stationary Rabi frequencies, given by $V_{j}=V_{j}\left(
\mathbf{r}\right)  =\mu_{j}\mathbf{E}_{j}(\mathbf{r})/\hbar$, where $\mu_{j}$
is the transition dipole moment. The complete set of Bloch equations for the
four-level \textit{N} system consists of sixteen equations
\cite{GorenN2004PRA}. In order to simplify the application of the theory to
EIA, we assume that $V_{p}\ll V_{1,2}<\Gamma$ and that the pump transitions
are well below saturation, so that in the absence of the probe, the population
concentrates in the $g_{2}$ state, the $g_{2} \leftrightarrow e_{2}$ dipole is
excited, and the $e_{2}$ state is empty up to second order in the pump field
\cite{GorenN2004PRA}. The equations can then be written up to the first order
in the probe field $V_{p}$ \cite{Taichenachev1999PRA}, which reduces the
number of Bloch equations to five.

The complete analytical development is presented in Appendix A, and an example
of the calculated probe absorption spectrum for collinear and degenerate beams
($\mathbf{q}_{1}=\mathbf{q}_{2}=\mathbf{q}$) is given in Fig. \ref{Fig. 2}(a)
(blue line). For the numerical calculations, we have considered the $D_{2}$
line of $^{85}$Rb (wavelength 780 nm) at room temperature, with a total
spontaneous emission rate $\Gamma=2\pi\times6$ MHz \cite{Lezama1999PRA}. Other
parameters are indicated in the figure caption and described in what follows.

Four complex frequencies control the EIA dynamics, each relating to a
different coherence in the process:%

\begin{subequations}
\label{Eq. 3}%
\begin{align}
\xi_{1}  &  =\left(  \Delta_{p}-\Delta_{1}\right)  -(\mathbf{q}_{p}%
-\mathbf{q}_{1})\cdot\mathbf{v+}i(\gamma+\gamma_{\text{vcc}}),\label{Eq. 3a}\\
\xi_{2}  &  =\Delta_{p}-\mathbf{q}_{p}\cdot\mathbf{v+}i(\tilde{\Gamma}%
+\gamma_{\text{vcc}}),\label{Eq. 3b}\\
\xi_{3}  &  =\left(  \Delta_{p}-\Delta_{2}\right)  -(\mathbf{q}_{p}%
-\mathbf{q}_{2})\cdot\mathbf{v+}i(\Gamma+\gamma+\gamma_{\text{vcc}%
}),\label{Eq. 3c}\\
\xi_{4}  &  =\left(  \Delta_{p}-\Delta_{1}-\Delta_{2}\right)  -(\mathbf{q}%
_{p}-\mathbf{q}_{1}-\mathbf{q}_{2})\cdot\mathbf{v+}i(\tilde{\Gamma}%
+\gamma_{\text{vcc}}), \label{Eq. 3d}%
\end{align}
with the one-photon detunings $\Delta_{j}=\omega_{j}-\omega_{e_{j}g_{j}}$
($j=1,2$) and $\Delta_{p}=\omega_{p}-\omega_{e_{1}g_{2}}$, and $\tilde{\Gamma
}=\Gamma/2+\Gamma_{\text{pcc}}+\gamma$. The frequency $\xi_{2}$ is related to
the probe transition and includes the one-photon Doppler shift $\mathbf{q}%
_{p}\cdot\mathbf{v}$. $\xi_{1}$ and $\xi_{3}$ relate to the slowly varying
ground and excited state coherences and include the residual Doppler shift
$(\mathbf{q}_{p}-\mathbf{q}_{i})\cdot\mathbf{v}$ and the Raman (two-photon)
detuning. $\xi_{4}$ relates to the three-photon transition (whose direct
optical-dipole is forbidden), required for the EIA process. Note that the fast
optical decay rates ($\Gamma$ or $\tilde{\Gamma}$) is absent only from
$\xi_{1}$.

In EIA, in contrast to EIT, a strong optical-dipole transition ($g_{2}%
\leftrightarrow e_{2}$) is excited even in the absence of the probe. Its
excitation depends on its resonance with the pump field, and is thus affected
by Doppler broadening. This leads to velocity-dependent equations even in
zero-order in the probe field, and introduces the additional complex frequency%
\end{subequations}
\begin{equation}
\xi_{5}=-\Delta_{2}+\mathbf{q}_{2}\cdot\mathbf{v+}i(\tilde{\Gamma}%
+\gamma_{\text{vcc}}),
\end{equation}
with the one-photon Doppler shift $\mathbf{q}_{2}\cdot\mathbf{v}$. The
overall dynamics is thus governed by the five equations (\ref{Eq. A11a}%
)-(\ref{Eq. A11e}).

We start by calculating the probe absorption spectrum for uniform pump and
probe fields (plane waves) by solving the equations analytically. The spectrum
depends on $18$ different integrals over velocity, of the form%

\begin{equation}
G_{i}=\int d^{3}v\frac{\xi_{\alpha}\cdots\xi_{\beta}}{\xi_{5}\xi_{d}%
}F(\mathbf{v}), \label{Gs}%
\end{equation}
where $F(\mathbf{v})=\left(  2\pi v_{\text{th}}^{2}\right)  ^{-3/2}%
e^{-\mathbf{v}^{2}/2v_{\text{th}}^{2}}$ is the Boltzmann velocity
distribution, and $v_{\text{th}}^{2}=k_{b}T/m$ is the mean thermal velocity.
The determinant $\xi_{d}$,
\begin{equation}
\xi_{d}=\xi_{1}\xi_{2}\xi_{3}\xi_{4}-\xi_{3}(\xi_{2}V_{2}^{2}+\xi_{4}V_{1}%
^{2})+iV_{1}V_{2}bA\Gamma\left(  \xi_{2}+\xi_{4}\right)  , \label{xie_d}%
\end{equation}
introduces the power broadening effect (first and second terms), \emph{i.e.}
the dependence of the Raman spectral width on the pump powers, and the
spontaneous TOC from the excited state to the ground state (last term). The
last term is associated with the TOC due to its dependence on the parameter
$b$, which sets the amount of TOC in the original dynamic equations
(\ref{Eq. A1}), and can take either the value 0 (no TOC) or 1
\cite{Taichenachev1999PRA}. The spontaneous decay branching ratio is given by
$A^{2}=\mu_{e_{1}g_{1}}^{2}/\left(  \mu_{e_{1}g_{1}}^{2}+\mu_{e_{1}g_{2}}%
^{2}\right)  $ \cite{GorenN2004PRA}. The TOC term in Eq. (\ref{xie_d}) depends
on the complex frequency%

\begin{align}
\xi_{2}+\xi_{4}  &  =\left(  2\Delta_{p}-\Delta_{1}-\Delta_{2}\right)
-(2\mathbf{q}_{p}-\mathbf{q}_{1}-\mathbf{q}_{2})\cdot\mathbf{v}\nonumber\\
&  \mathbf{+}2i(\Gamma/2+\Gamma_{\text{pcc}}+\gamma+\gamma_{\text{vcc}}).
\label{Xi2Xi4}%
\end{align}
It is important to note that, although each of the individual frequencies
$\xi_{2}$ and $\xi_{4}$ is affected by a Doppler shift (either one- or
three-photon), the sum $\xi_{2}+\xi_{4}$ exhibits \emph{only a residual
Doppler shift} (assuming nearly collinear pumps, $\mathbf{q}_{1}%
\approx\mathbf{q}_{2}$). Nevertheless the relaxation rate $(\Gamma
/2+\Gamma_{\text{pcc}}+\gamma+\gamma_{\text{vcc}})$ is the same as that
characterizing the decay of the optical transitions. As a consequence, even
when $\Gamma_{\text{pcc}}$ is much smaller than the optical Doppler width, it
plays a significant role in determining the intensity of the EIA spectrum.
This is in contrast to one- and two-photon processes (such as EIT), in which
$\Gamma_{\text{pcc}}$ is irrelevant when it is much smaller than the Doppler
width. It can also be seen that when $\mathbf{q}_{1}\approx\mathbf{q}_{2},$
the various residual Doppler shifts are negligible compared to the relaxation
rates in the determinant $\xi_{d},$ so that $\xi_{d}$ is only weakly dependent
on these shifts.

Examining the absorption spectrum in Fig. \ref{Fig. 2}(a), we observe the
narrow absorption peak on top of the broad one-photon curve. Moreover, as can
be seen in the inset, the EIA resonance consists of two independent features:
a ``pedestal" at the base and a sharp absorption peak at the center. In order
to obtain physical insight into these features, we have derived an approximate
solution for the probe absorption which incorporates the main contributions to
the EIA, namely the underlying EIT mechanism plus the spontaneous TOC. The
approximate Fourier transform of the nondiagonal density-matrix element for
the probe is%

\begin{equation}
R_{e_{1}g_{2}}=n_{0}\left[  -G_{4}+V_{2}^{2}G_{5}+iV_{1}V_{2}bA\Gamma
\frac{iG_{2}G_{3}\gamma_{\text{vcc}}}{1-iG_{1}\gamma_{\text{vcc}}}\right]
V_{p}, \label{Eq. 2}%
\end{equation}
where $G_{1}=\int d^{3}v\frac{\xi_{2}\xi_{3}\xi_{4}F(\mathbf{v})}{\xi_{d}}$,
$G_{2}=\int d^{3}v\frac{\xi_{3}\xi_{4}F(\mathbf{v})}{\xi_{d}}$, $G_{3}=\int
d^{3}v\frac{\xi_{2}\xi_{4}F(\mathbf{v})}{\xi_{5}\xi_{d}}$, $G_{4}=\int
d^{3}v\frac{\xi_{1}\xi_{3}\xi_{4}F(\mathbf{v})}{\xi_{d}}$, $G_{5}=\int
d^{3}v\frac{\xi_{3}F(\mathbf{v})}{\xi_{d}}$, and $n_{0}$ is the number density
of the active atoms. It can be shown that Eq. (\ref{Eq. 2}) is valid provided
$\gamma_{\text{vcc}}\ll\Gamma_{\text{pcc}}+\Gamma/2.$ For an atom at rest and
in the absence of collisions, so that $\gamma_{\text{vcc}}=0,$ $v_{\text{th}%
}\rightarrow0,$ and $\Gamma_{\text{pcc}}=0,$ Eq. (\ref{Eq. 2}) is identical to
the expression obtained by Taichenachev \textit{et al}.
\cite{Taichenachev1999PRA} (with $b=1$),
\begin{equation}
R_{e_{1}g_{2}}^{\text{rest}}=\frac{in_{0}V_{p}}{\Gamma/2-i\Delta_{p}}\left[
1+\frac{2A\left\vert V_{1}\right\vert ^{2}/\Gamma}{2\left(  1-A^{2}\right)
\left\vert V_{2}\right\vert ^{2}/\Gamma-i\Delta_{p}}\right]  . \label{Eq. 2a}%
\end{equation}
The first term in the square brackets in Eqs. (\ref{Eq. 2}) and (\ref{Eq. 2a})
describes the one-photon (background) absorption, and the other terms are the
EIA peak.

For a moving atom, the spectrum resulting from Eq. (\ref{Eq. 2}) is shown in
Fig. \ref{Fig. 2}(a) (red dashed line) and is compared with the exact
solution; evidently, there is a good agreement between the spectra. Despite
the small discrepancy in the intensity of the sharp peak, the approximate
solution preserves the main features in the resonance. When plotted separately
in Fig. \ref{Fig. 2}(b), the three terms in Eq. (\ref{Eq. 2}) can be
identified with the different spectral features: $-G_{4}$ (black dashed line)
describes the background absorption; $V_{2}^{2}G_{5}$ (brown dotted line),
which constitutes the total peak in the absence of VCC, describes the wide
pedestal; and $iG_{2}G_{3}\gamma_{\text{vcc}}/(1-G_{1}\gamma_{\text{vcc}})$
(green dashed-dotted line) describes the sharp EIA peak, induced by VCC.%
\begin{figure}
[ptb]
\begin{center}
\includegraphics[
height=2.5642in,
width=3.4696in
]%
{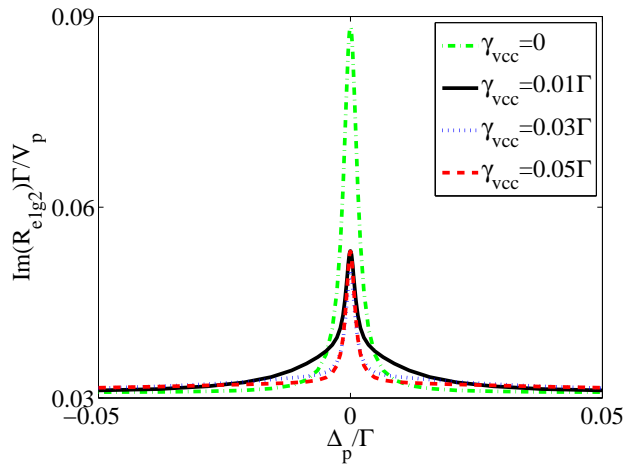}%
\caption{The EIA\ peak for different $\gamma_{\text{vcc}}$ rates and
$\Gamma_{\text{pcc}}=5\Gamma$ at zero pump-probe angular deviation; other
parameters as in Fig. \ref{Fig. 2} }%
\label{Fig. 3}%
\end{center}
\end{figure}

Fig. \ref{Fig. 3} shows the effect of varying the VCC rate, for a fixed PCC
rate ($\Gamma_{\text{pcc}}=5\Gamma$) and zero pump-probe angular deviation.
The width of the pedestal feature depends on the VCC rate and is given by
$\gamma_{\text{vcc}}+\gamma,$ while the width of the narrow peak shows only a
very weak dependence on $\gamma_{\text{vcc}}$. Increasing the VCC rate leads
to a decrease in the overall EIA\ intensity, but to an increase in the ratio
between the amplitude of the narrow peak and the pedestal baseline.

\begin{figure}
[ptb]
\begin{center}
\includegraphics[
height=2.5222in,
width=3.4686in
]%
{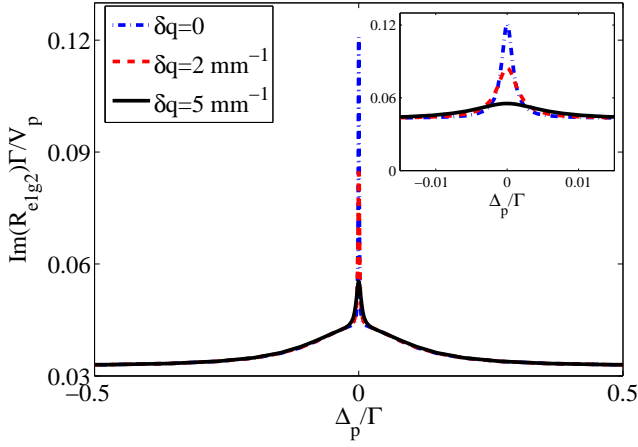}%
\caption{Calculated probe absorption spectra with $\gamma_{\text{vcc}%
}=0.1\Gamma$ and $\Gamma_{\text{pcc}}=\Gamma$, for different pump-probe
angular deviations. Other parameters as in Fig. \ref{Fig. 2}.}%
\label{Fig. 4}%
\end{center}
\end{figure}

\begin{figure}
[ptb]
\begin{center}
\includegraphics[
height=3.0261in,
width=3.4686in
]%
{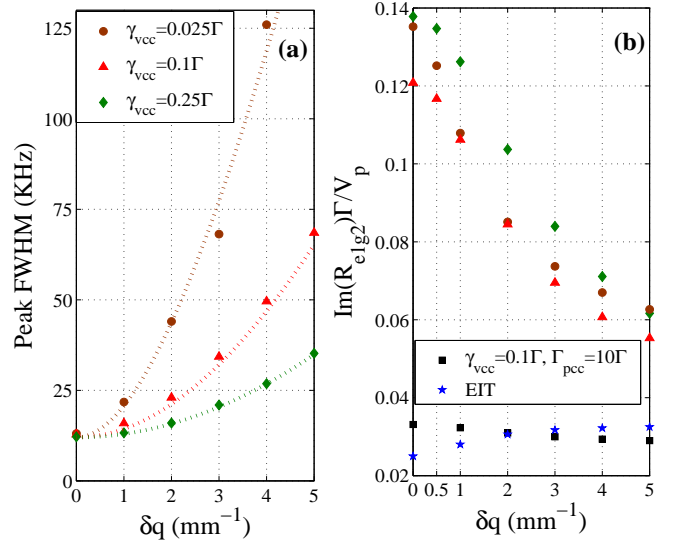}%
\caption{Calculated EIA FWHM (a) and absorption (b) for $\Gamma_{\text{pcc}%
}=\Gamma$ and various values of $\gamma_{\text{vcc}}$, as a function of the
pump-probe wave-vector difference $\delta\mathbf{q}$. The blue starts are the
EIT absorption on the Raman resonance conditions, with $\Gamma_{\text{pcc}%
}=\Gamma$ and $\gamma_{\text{vcc}}=0.025\Gamma$.}%
\label{Fig. 5}%
\end{center}
\end{figure}

We now turn to explore the residual (two-photon and four-photon) Doppler and
Dicke effects due to wave-vector mismatch between the pump fields and the
probe, introduced in principle either by a frequency detuning between the
fields, $|\mathbf{q}_{p}|\neq|\mathbf{q}_{1,2}|,$ or due to an angular
deviation between them, $\mathbf{q}_{p}\nparallel\mathbf{q}_{1,2}$. We mainly
focus on the latter, which may be found in a nearly degenerate level scheme,
and we further take the two pump fields to be the same, namely $\mathbf{q}%
_{1}=\mathbf{q}_{2}$. Figure \ref{Fig. 4} presents the probe absorption
spectrum for different values of the wave-vector difference, $\delta
\mathbf{q=}$ $\mathbf{q}_{p}-\mathbf{q}_{1,2},$ when $\gamma_{\text{vcc}%
}=0.1\Gamma$ and $\Gamma_{\text{pcc}}=\Gamma$. As can be seen, increasing
$\delta\mathbf{q}$ broadens the EIA spectrum (see inset). This is analogous to
the broadening of an EIT transmission peak in a similar configuration
\cite{Shuker2007PRA}. However, the wide collisionally-broadened pedestal
remains unaffected by the changes in $\delta\mathbf{q}$, indicating that it
mostly originates from homogenous decay processes. Figure \ref{Fig. 5}(a)
summarizes the full-width at half-maximum (FWHM) of the EIA peak for
$\Gamma_{\text{pcc}}=\Gamma$ and for various values of $\gamma_{\text{vcc}}$,
as a function of $\delta\mathbf{q}$. Because of the difficulty of separating
the sharp peak from the background in the calculated spectra \footnote{When
$\gamma_{\text{vcc}}\rightarrow0$, the third term in Eq. (\ref{Eq. 2})
gradually vanishes, and the second term ($V_{2}^{2}G_{5}$) is responsible for
the EIA peak, as indicated by the brown-dotted line in Fig. \ref{Fig. 2}(b).
Its width is limited by homogenous broadening mechanisms and determined by
$\gamma$ and $\Gamma_{\text{pcc}}$.}, the widths of the sharp EIA peak were
obtained only from the third term in Eq. (\ref{Eq. 2}). In contrast to an EIT
peak, which does not depend on $\gamma_{\text{vcc}}$ when $\delta\mathbf{q=0}$
(collinear degenerate beams) \cite{Firstenberg2008PRA}, the FWHM of the EIA
peak at $\delta\mathbf{q=0}$ depends weakly on the VCC rate (although barely
noticeable in the figure). This difference derives from the effect of
collisions on the pump absorption in the case of EIA, as described earlier.%

For $\delta\mathbf{q\neq0}$ the FWHM of the peak in the Dicke limit (high
$\gamma_{\text{vcc}}$) depends on $\gamma_{\text{vcc}}$ and is proportional to
the residual Doppler-Dicke width, $2v_{\text{th}}\delta\mathbf{q}^{2}%
/\gamma_{\text{vcc}}$. In this limit, the results are well approximated by the
analytic expression \cite{Firstenberg2008PRA} [dotted lines in Fig.
\ref{Fig. 5}(a)]:%

\begin{equation}
\text{FWHM}=2\times\frac{2}{a^{2}}\gamma_{\text{vcc}}H\left(  a\frac
{v_{\text{th}}\delta q}{\gamma_{\text{vcc}}}\right)  , \label{Eq. 5}%
\end{equation}
where $H\left(  x\right)  =e^{-x}-1+x$ and $a^{2}=2/\ln2$. Increasing the
pump-probe angular deviation reduces the efficiency of the EIA process and
thus results in a decrease in the probe absorption [Fig. \ref{Fig. 5}(b)].
This is of course the opposite trend to that of EIT (blue stars), where the
depth of the dip decreases (the absorption increases) with increasing $\delta
q$ \cite{Firstenberg2008PRA}.

\section{\label{sec:filter}Spatial-frequency filter}

We now to turn to discuss the results of our model from the viewpoint of a
spatial-frequency filter for a structured probe beam. When non-uniform beams
are considered, the different spatial frequencies that comprise the beams
result in different Doppler and Dicke widths. Consequently, the various
spatial-frequency components experience different absorption and refraction in
the medium. Specifically, the dependence of the absorption on the transverse
wave-vectors of the probe beam manifests a filter for the probe in Fourier space.

We assume an optical configuration of two collinear uniform pumps (plane waves
with $V_{1}$ and $V_{2}$ constant) and a spatially varying propagating probe,
$V_{p}=V_{p}(\mathbf{r},t)$. Since the medium exhibits a non-local response
due the atomic motion, the evolution of the probe is more naturally described
in the Fourier space $V_{p}(\mathbf{k},\omega)$ where $\mathbf{k}$ and
$\omega$ are the spatial and temporal frequencies of the envelope of the
probe. Under these assumptions, the model results in a Diffusion-like
equations for the populations and coherences of the atomic medium, derived in
Appendix B. To simplify the general dynamics of Eqs. (\ref{Eq. B8a}) and
(\ref{Eq. B9}), we take the stationary case [$\omega=0,$ $V_{p}=V_{p}%
(\mathbf{k})$] and assume that the carrier wave-vector of the probe is the
same as that of the pumps, $\mathbf{q}_{p}=\mathbf{q}_{1}=\mathbf{q}_{2},$ so
that $\delta\mathbf{q}_{1}=\delta\mathbf{q}_{2}=0$. Taking the Fourier
transform [see Eq. (\ref{Eq. A13})], we obtain a set of steady-state equations
for the spatially-dependent atomic coherences, $R_{g_{1}g_{2}}(\mathbf{k})$,
$R_{e_{1}e_{2}}(\mathbf{k})$, and $R_{e_{1}g_{2}}(\mathbf{k}),$
\begin{subequations}
\label{Eq. 6}%
\begin{align}
&  \left[  i\left(  \Delta_{p}-\Delta_{1}\right)  -\gamma-K_{\text{1p}%
}\left\vert V_{1}\right\vert ^{2}-K_{\text{3p}}\left\vert V_{2}\right\vert
^{2}-Dk^{2}\right]  R_{g_{1}g_{2}}\nonumber\\
&  =(Dk^{2}-bA\Gamma)R_{e_{1}e_{2}}+K_{\text{1p}}V_{1}^{\ast}V_{p}%
n_{0},\label{Eq. 6a}\\
&  \left[  i\left(  \Delta_{p}-\Delta_{2}\right)  -\Gamma-\gamma
-Dk^{2}\right]  R_{e_{1}e_{2}}\nonumber\\
&  =-V_{1}V_{2}^{\ast}(K_{\text{1p}}+K_{\text{3p}})R_{g_{1}g_{2}}-V_{2}^{\ast
}(K_{\text{1p}}+K_{\text{pump}})V_{p}n_{0},\label{Eq. 6b}\\
&  R_{e_{1}g_{2}}=iK_{\text{1p}}\left(  V_{1}R_{g_{1}g_{2}}+V_{p}n_{0}\right)
\label{Eq. 6c}%
\end{align}
where $K_{\text{1p}}=iG_{\text{1p}}/\left(  1-iG_{\text{1p}}\gamma
_{\text{vcc}}\right)  $ is the one-photon absorption spectrum with
$G_{\text{1p}}=\int F\left(  \mathbf{v}\right)  /\xi_{2}d^{3}v\mathbf{\ }$;
$K_{\text{3p}}=iG_{\text{3p}}/\left(  1-iG_{\text{3p}}\gamma_{\text{vcc}%
}\right)  $ is the three-photon absorption spectrum with $G_{\text{3p}}=\int
F\left(  \mathbf{v}\right)  /\xi_{4}d^{3}v;\mathbf{\ }$and $K_{\text{pump}%
}=iG_{\text{pump}}/\left(  1-iG_{\text{pump}}\gamma_{\text{vcc}}\right)  $ is
the one-photon (pump) absorption spectrum with $G_{\text{pump}}=\int F\left(
\mathbf{v}\right)  /\xi_{5}d^{3}\mathbf{v}$, as described in Appendix B.
Solving Eq. (\ref{Eq. 6}) for $R_{e_{1}g_{2}}\left(  \mathbf{k},\omega\right)
$, substituting the result into the expression for the linear-susceptibility
[Eq. (\ref{Eq. A16})], assuming that $V_{1}=\eta V_{2}$ ($0<\eta\leq1$), and
neglecting all the terms proportional to $1/\Gamma$, we obtain%

\end{subequations}
\begin{subequations}
\label{Eq. 7}%
\begin{align}
&  \chi_{e_{1}g_{2}}\left(  \mathbf{k}\right)  =\frac{g}{c}iKn_{0}\left(
1+\text{L}\right)  ,\label{Eq. 7a}\\
&  \text{L}=\frac{\eta\left(  2bA-\eta\right)  \Gamma_{\text{p}}}{-i\Delta
_{p}+\gamma+\left(  \eta^{2}+1-2bA\eta\right)  \Gamma_{\text{p}}+Dk^{2}%
},\label{Eq. 7b}%
\end{align}
where $D=v_{th}/\gamma_{\text{vcc}}$ is the diffusion coefficient,
$\Gamma_{\text{p}}=K\left\vert V_{2}\right\vert ^{2}$ is the power broadening,
and $K_{\text{1p}}\approx K_{\text{3p}}\approx K_{\text{pump}}=K=\int F\left(
\mathbf{v}\right)  /\left[  \mathbf{q}_{p}\cdot\mathbf{v+}i\left(
\Gamma/2+\Gamma_{\text{pcc}}+\gamma+\gamma_{\text{vcc}}\right)  \right]
d^{3}v$ for $\Delta_{p}\ll$ $\Gamma_{\hom}=\gamma+\Gamma_{\text{p}}$. In the
case where $\eta=A,$ Eqs. (\ref{Eq. 7}) is similar to Eq. (\ref{Eq. 2a})
obtained by Taichenachev \textit{et al. }\cite{Taichenachev1999PRA}, except
for the diffusion term $Dk^{2}$, which vanishes for an atom at rest.%
\begin{figure}
[ptb]
\begin{center}
\includegraphics[
height=2.5356in,
width=3.4705in
]%
{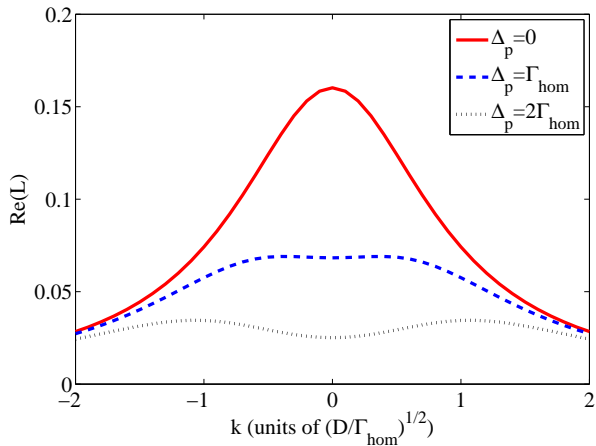}%
\caption{The EIA spatial-frequency filter, given in Eq. (\ref{Eq. 7b}), as a function of $k$, with
$\gamma_{\text{vcc}}=0.025\Gamma$ and $\Gamma _{\text{pcc}}=10\Gamma$. Red curve is plotted for $\Delta_{p}$ on
resonance, blue-dashed and black-dotted curves demonstrate the behavior at nonzero Raman
detuning.}%
\label{Fig. 6}%
\end{center}
\end{figure}

The imaginary part of the susceptibility in Eq. (\ref{Eq. 7}) yields the
absorption of the probe for various values of $\mathbf{k}$. The first term in
the brackets in Eq. (\ref{Eq. 7a}) is the linear one-photon absorption, and
the second term is the $k$-dependent EIA contribution. Thus, the real part of
$L$ in Eq. (\ref{Eq. 7b}) describes an \textquotedblleft
absorbing\textquotedblright\ spatial-frequency filter, the same way as was
done for EIT \cite{Shuker2008PRL,Firstenberg2009NP}. Fig. \ref{Fig. 6}
summarizes several examples of the EIA spatial filter behavior as a function
of $k$ for $\Delta_{p}=0,$ $\Delta_{p}=\pm\Gamma_{\hom},$ and $\Delta_{p}%
=\pm2\Gamma_{\hom}$. At $\Delta_{p}=0,$ the curve is a Lorentzian and maximum
absorption is achieved. When $\Delta_{p}\neq0$ the filter becomes more transparent.

\section{\label{sec:Ramseynarr}Ramsey narrowing}
We now consider the \textit{N} system interacting with collinear probe and pump beams that have finite widths.
Due to thermal motion, the alkali atoms spend a period of time in the interaction region and then leave the
light beams, evolve `in the dark', and diffuse back inside. Such a random periodic
motion was described recently by Xiao \textit{et al. }%
\cite{Xiao2006PRL,Xiao2008OE} for an EIT system, and was shown to result in a
cusp-like spectrum. Near its center, the line is much narrower than that
expected from time-of-flight broadening and power broadening, and the effect,
resulting from the contribution of bright-dark-bright atomic trajectories of
random durations, was named Ramsey narrowing.%

\begin{figure}
[ptb]
\begin{center}
\includegraphics[
height=2.6948in,
width=3.4705in
]%
{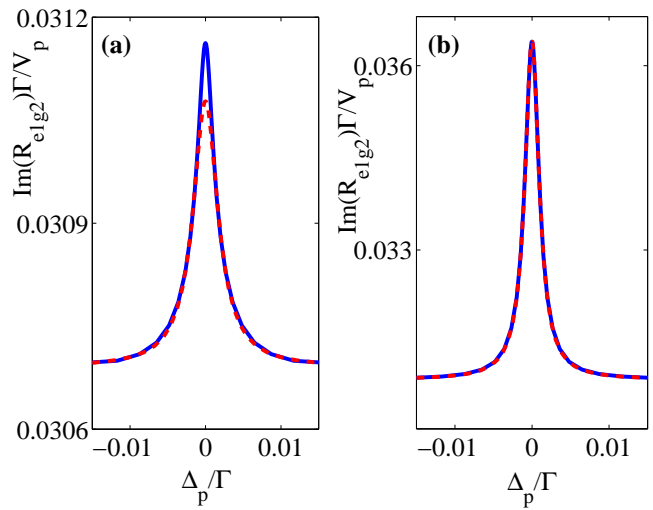}%
\caption{Calculated probe absorption spectra (blue line) for one-dimensional
stepwise beam with finite thickness: (a) $2a=100$ $\mu m$ and (b) $2a=10$
$mm$, and fitted Lorentzian (red dashed line). All other parameters are the
same as in Fig.\ref{Fig. 2}}%
\label{Fig. 7}%
\end{center}
\end{figure}

Ramsey-narrowed spectra can be calculated analytically from the diffusion
equations of the atomic coherences when the light fields of both the probe and
pump beams have finite widths \cite{Firstenberg2008PRA}. The EIA spectrum
resulting from a one-dimensional uniform light-sheet of thickness $2a$ in the
$x-$direction is derived analytically in Appendix B [Eq. (\ref{B15})]. In Fig.
\ref{Fig. 7}, we show the spectrum for two different thicknesses and the
fitted Lorentzian curves. Near the resonance, the EIA line for the $100$ $\mu
m$ sheet is spectrally sharper than the fitted Lorentzian -- the
characteristic signature of Ramsey narrowing. In contrast, the EIA peak
calculated for a $10$ $mm$ beam is well fitted by the Lorentzian. In addition,
the EIA contrast deteriorates as the beam becomes narrower, since the
interaction area decreases and fewer atoms interact with the fields.

\section{\label{sec:conc}Conclusions}

In this paper, we extended the theory that describes the effect of buffer-gas
collisions on three-level $\Lambda$ systems in an EIT configuration
\cite{Firstenberg2007PRA,Shuker2007PRA,Firstenberg2008PRA} to the case of a
four-level closed $N$ system which is the simplest system that experiences EIA
due to TOC. Using this formalism, we investigated the influence of collisions
of optically active atoms with a buffer gas on the EIA peak. In addition to
the exact expressions, we presented an approximate solution for the probe
absorption spectrum, which provides a physical insight into the behavior of
the EIA peak due to VCC, PCC, and wave-vector difference between the pump and
probe beams. VCC were shown to produce a wide pedestal at the base of the
EIA\ peak; increasing the pump-probe angular deviation scarcely affects the
pedestal whereas the sharp central EIA peak becomes weaker and broader due to
the residual Doppler-Dicke effect. Using diffusion-like equations for the
atomic coherences and populations, the spatial-frequency filter and the
Ramsey-narrowed spectrum were analytically obtained.

In extending the description from the $\Lambda$ to the $N$ schemes, we have
considered several elements that are likely to be important in other
four-level systems. These include the diffusion of excited-state coherences
and the influence of the thermal motion on the optical dipole in the absence
of the probe. The latter introduces a Doppler contribution into the pumping
terms and consequently affects the power broadening of the narrow resonances.
\end{subequations}
\appendix

\section{Reduced density matrix}

Consider the near-resonant interaction of a light field consisting of one or
two moderately strong pumps and a weak probe, as given in Eq. (\ref{Eq. 1}),
with the four-level degenerate $N$\textit{ }system of Fig. \ref{Fig. 1}(a). We
use the first-order approximation in the probe amplitude, $V_{p}$, and assume
that $V_{2}<\Gamma,$ $V_{1}\leqslant V_{2},$ $V_{p}<V_{1,2}$. Since the pump
transitions are assumed non-saturated, the atomic population in the absence of
the probe concentrates in the $g_{2}$ state, and the population in other
states can be neglected. The $g_{2} \leftrightarrow e_{2}$ dipole, excited in
the absence of the probe, is of importance and is thus considered. The
resulting Bloch equations are \cite{Taichenachev1999PRA}


\begin{subequations}
\label{Eq. A1}%
\begin{align}
\dot{\breve{\rho}}_{g_{1}g_{2}}^{\left(  1\right)  ,i}  &  \left(  \omega
_{p}-\omega_{1}\right)  =-\left[  i\left(  \omega_{e_{1}g_{2}}-\omega
_{e_{1}g_{1}}\right)  +\gamma\right]  \breve{\rho}_{g_{1}g_{1}}^{\left(
1\right)  ,i}\nonumber\\
&  +i\breve{V}_{1}^{\ast}\breve{\rho}_{e_{1}g_{2}}^{\left(  1\right)
,i}-i\breve{V}_{2}\breve{\rho}_{g_{1}e_{2}}^{\left(  1\right)  ,i}%
+bA\Gamma\breve{\rho}_{e_{1}e_{2}}^{\left(  1\right)  ,i},\label{Eq. A1a}\\
\dot{\breve{\rho}}_{e_{1}g_{2}}^{\left(  1\right)  ,i}  &  \left(  \omega
_{p}\right)  =-\left[  i\omega_{e_{1}g_{2}}+\Gamma/2+\Gamma_{\text{pcc}%
}\right]  \breve{\rho}_{e_{1}g_{2}}^{\left(  1\right)  ,i}\nonumber\\
&  +i\breve{V}_{p}\breve{\rho}_{g_{2}g_{2}}^{\left(  0\right)  ,i}+i\breve
{V}_{1}\breve{\rho}_{g_{1}g_{2}}^{\left(  1\right)  ,i},\label{Eq. A1b}\\
\dot{\breve{\rho}}_{e_{1}e_{2}}^{\left(  1\right)  ,i}  &  \left(  \omega
_{p}-\omega_{2}\right)  =-\left[  i\left(  \omega_{e_{1}g_{2}}-\omega
_{e_{2}g_{2}}\right)  +\Gamma+\gamma\right]  \breve{\rho}_{e_{1}e_{2}%
}^{\left(  1\right)  ,i}\nonumber\\
&  +i\breve{V}_{p}\breve{\rho}_{g_{2}e_{2}}^{\left(  0\right)  ,i}+i\breve
{V}_{1}\breve{\rho}_{g_{1}e_{2}}^{\left(  1\right)  ,i}-i\breve{V}_{2}^{\ast
}\breve{\rho}_{e_{1}g_{2}}^{\left(  1\right)  ,i},\label{Eq. A1c}\\
\dot{\breve{\rho}}_{g_{1}e_{2}}^{\left(  1\right)  ,i}  &  \left(  \omega
_{p}-\omega_{1}-\omega_{2}\right)  =-\left[  i\left(  \omega_{e_{1}g_{2}%
}-\omega_{e_{1}g_{1}}-\omega_{e_{2}g_{2}}\right)  \right. \nonumber\\
&  +\Gamma/2+\left.  \Gamma_{\text{pcc}}\right]  \breve{\rho}_{g_{1}g_{1}%
}^{\left(  1\right)  ,i}-i\breve{V}_{2}^{\ast}\breve{\rho}_{g_{1}g_{2}%
}^{\left(  1\right)  ,i},\label{Eq. A1d}\\
\dot{\breve{\rho}}_{g_{2}e_{2}}^{\left(  0\right)  ,i}  &  \left(  -\omega
_{2}\right)  =-\left[  \Gamma/2+\Gamma_{\text{pcc}}-i\omega_{e_{2}g_{2}%
}\right]  \breve{\rho}_{g_{2}e_{2}}^{\left(  0\right)  ,i}\left(  -\omega
_{2}\right) \nonumber\\
&  +i\breve{V}_{2}^{\ast}\left(  \breve{\rho}_{e_{2}e_{2}}^{\left(  0\right)
,i}-\breve{\rho}_{g_{2}g_{2}}^{\left(  0\right)  ,i}\right)  . \label{Eq. A1e}%
\end{align}
Here, $\breve{\rho}_{ss^{\prime}}^{\left(  j\right)  ,i}$ is the
density-matrix element of the $i-$th atom (one of many identical particles) to
the $j-$th order in the probe, and apart from $\breve{\rho}_{g_{2}g_{2}%
}^{\left(  0\right)  ,i}\approx1,$ $\breve{\rho}_{ss}^{\left(  0\right)
,i}=0$. We also consider the envelopes of the pumps to be constant in time so
that $V_{1,2}$ is shorthand for $V_{1,2}\left(  \mathbf{r}\right)  $. The wave
equation for the probe field is%

\end{subequations}
\begin{equation}
\left(  \nabla^{2}-\frac{1}{c^{2}}\frac{\partial^{2}}{\partial t^{2}}\right)
\mathbf{\breve{E}}_{p}\left(  \mathbf{r},t\right)  =\frac{4\pi}{c^{2}}%
\frac{\partial^{2}}{\partial t^{2}}\mathbf{\breve{P}}_{e_{1}g_{2}}\left(
\mathbf{r},t\right)  , \label{Eq. A2}%
\end{equation}
where $\mathbf{\breve{P}}_{e_{1}g_{2}}\left(  \mathbf{r},t\right)
=\mathbf{P}_{e_{1}g_{2}}\left(  \mathbf{r},t\right)  e^{-i\omega_{p}%
t}e^{-i\mathbf{q}_{p}\cdot t}$ is the contribution of the $e_{1}%
\leftrightarrow g_{2}$ transition to the expectation value of the
polarization, $\mathbf{P}_{e_{1}g_{2}}$ is the slowly varying polarization,
and $\nabla^{2}$ is the three-dimensional Laplacian operator. With Eq.
(\ref{Eq. 1}), and assuming without loss of generality that $\mathbf{\hat{q}%
}_{p}=\mathbf{\hat{z}}q_{p}$, as shown in Fig. \ref{Fig. 1}(b), Eq.
(\ref{Eq. A2}) can be written in the paraxial approximation as%

\begin{equation}
\left(  \frac{\partial}{\partial t}+c\frac{\partial}{\partial z}-i\frac
{c}{2q_{p}}\nabla_{\bot}^{2}\right)  V_{p}\left(  \mathbf{r},t\right)
=i\frac{g}{\mu_{e_{1}g_{2}}^{\ast}}\mathbf{P}_{e_{1}g_{2}}\left(
\mathbf{r},t\right)  , \label{Eq. A3}%
\end{equation}
where $\nabla_{\bot}^{2}$ is the transverse Laplacian operator, and
$g=2\pi\omega_{p}\left\vert \mu_{e_{1}g_{2}}\right\vert ^{2}/\hbar$ is a
coupling constant .

Following \cite{Firstenberg2008PRA}, we introduce a density-matrix
distribution function in space and velocity,%

\begin{equation}
\breve{\rho}_{ss^{\prime}}^{{}}=\breve{\rho}_{ss^{\prime}}^{{}}\left(
\mathbf{r},\mathbf{v},t\right)  =\underset{i}{\sum}\breve{\rho}_{ss^{\prime}%
}^{i}\left(  t\right)  \delta\left(  \mathbf{r-r}_{i}\left(  t\right)
\right)  \delta\left(  \mathbf{v-v}_{i}\left(  t\right)  \right)  ,
\label{Eq. A4}%
\end{equation}
where the time dependence of $\breve{\rho}_{ss^{\prime}}^{i}\left(  t\right)
$ is determined by Eqs. (\ref{Eq. A1}). Differentiating Eq. (\ref{Eq. A4})
with respect to time, we arrive at
\begin{align}
&  \frac{\partial}{\partial t}\breve{\rho}_{ss^{\prime}}^{{}}+\mathbf{v}%
\cdot\frac{\partial}{\partial\mathbf{r}}\breve{\rho}_{ss^{\prime}}^{{}%
}+\left[  \frac{\partial}{\partial t}\breve{\rho}_{ss^{\prime}}^{{}}\right]
_{\operatorname{col}}\nonumber\\
&  =\underset{i}{\sum}\frac{\partial}{\partial t}\breve{\rho}_{ss^{\prime}%
}^{i}\left(  t\right)  \delta\left(  \mathbf{r-r}_{i}\left(  t\right)
\right)  \delta\left(  \mathbf{v-v}_{i}\left(  t\right)  \right)  ,
\label{Eq. A5}%
\end{align}
where the effect of velocity-changing collisions is taken in the strong
collision limit in the form of a Boltzmann relaxation term
\cite{Sobelman1967SPU},%

\begin{equation}
\left[  \frac{\partial}{\partial t}\breve{\rho}_{ss^{\prime}}^{{}}\right]
_{\operatorname{col}}=-\gamma_{\text{vcc}}\left[  \breve{\rho}_{ss^{\prime}%
}^{{}}\left(  \mathbf{r},\mathbf{v},t\right)  -\breve{R}_{ss^{\prime}}^{{}%
}\left(  \mathbf{r},t\right)  F(\mathbf{v})\right]  , \label{Eq. A6}%
\end{equation}
with $\breve{R}_{ss^{\prime}}^{{}}=\breve{R}_{ss^{\prime}}^{{}}\left(
\mathbf{r},t\right)  =\int d^{3}\mathbf{v}\breve{\rho}_{ss^{\prime}}^{{}%
}\left(  \mathbf{r},\mathbf{v},t\right)  $ being the density-number of atoms
per unit volume, near $\mathbf{r}$ in space, and%
\begin{equation}
F=F(\mathbf{v})=\left(  2\pi v_{\text{th}}\right)  ^{-3/2}e^{-\mathbf{v}%
^{2}/2v_{\text{th}}},\text{ }v_{\text{th}}=\frac{k_{b}T}{m}%
\end{equation}
is the Boltzmann distribution.

Before writing the coupled dynamics of the internal and motional degrees of
freedom, we introduce the slowly varying envelopes of the density-matrix
elements, $\rho_{ss^{\prime}}=\rho_{ss^{\prime}}\left(  \mathbf{r}%
,\mathbf{v},t\right)  $, as%
\begin{align}
&  \breve{\rho}_{g_{1}g_{2}}=\rho_{g_{1}g_{2}}e^{-i\left(  \omega_{p}%
-\omega_{1}\right)  t}e^{i\left(  \mathbf{q}_{p}-\mathbf{q}_{1}\right)
\cdot\mathbf{r}},\nonumber\\
&  \breve{\rho}_{e_{1}g_{2}}=\rho_{e_{1}g_{2}}e^{-i\omega_{p}t}e^{i\mathbf{q}%
_{p}\cdot\mathbf{r}},\nonumber\\
&  \breve{\rho}_{e_{1}e_{2}}=\rho_{e_{1}e_{2}}e^{i\left(  \mathbf{q}%
_{p}-\mathbf{q}_{2}\right)  \cdot\mathbf{r}},\nonumber\\
&  \breve{\rho}_{g_{1}e_{2}}=\rho_{g_{1}e_{2}}e^{-i\left(  \omega_{p}%
-\omega_{1}-\omega_{2}\right)  t}e^{i\left(  \mathbf{q}_{p}-\mathbf{q}%
_{1}-\mathbf{q}_{2}\right)  \cdot\mathbf{r}},\nonumber\\
&  \breve{\rho}_{g_{2}e_{2}}=\rho_{g_{2}e_{2}}e^{i\omega_{2}t}e^{-i\mathbf{q}%
_{2}\cdot\mathbf{r}}, \label{Eq. A9}%
\end{align}
and similarly the slowly varying densities $R_{ss^{\prime}}^{{}}=\int
d^{3}\mathbf{v}\rho_{ss^{\prime}}^{{}}$. Eqs. (\ref{Eq. A1}) then become
\begin{subequations}
\label{Eq. A11}%
\begin{align}
&  \left[  \frac{\partial}{\partial t}+\mathbf{v}\cdot\frac{\partial}%
{\partial\mathbf{r}}-i\xi_{1}\right]  \rho_{g_{1}g_{2}}-\gamma_{\text{vcc}%
}R_{g_{1}g_{2}}F\nonumber\\
&  =i\left(  V_{1}^{\ast}\rho_{e_{1}g_{2}}-V_{2}\rho_{g_{1}e_{2}}\right)
+bA\Gamma\rho_{e_{1}e_{2}},\label{Eq. A11a}\\
&  \left[  \frac{\partial}{\partial t}+\mathbf{v}\cdot\frac{\partial}%
{\partial\mathbf{r}}-i\xi_{2}\right]  \rho_{e_{1}g_{2}}-\gamma_{\text{vcc}%
}R_{e_{1}g_{2}}F\nonumber\\
&  =i\left[  V_{p}n_{0}F+V_{1}\rho_{g_{1}g_{2}}\right]  ,\label{Eq. A11b}\\
&  \left[  \frac{\partial}{\partial t}+\mathbf{v}\cdot\frac{\partial}%
{\partial\mathbf{r}}-i\xi_{3}\right]  \rho_{e_{1}e_{2}}-\gamma_{\text{vcc}%
}R_{e_{1}e_{2}}F\nonumber\\
&  =i\left(  V_{1}\rho_{g_{1}e_{2}}-V_{2}^{\ast}\rho_{e_{1}g_{2}}\right)
+iV_{p}^{{}}\rho_{g_{2}e_{2}},\label{Eq. A11c}\\
&  \left[  \frac{\partial}{\partial t}+\mathbf{v}\cdot\frac{\partial}%
{\partial\mathbf{r}}-i\xi_{4}\right]  \rho_{g_{1}e_{2}}-\gamma_{\text{vcc}%
}R_{g_{1}e_{2}}F\nonumber\\
&  =-iV_{2}^{\ast}\rho_{g_{1}g_{2}},\label{Eq. A11d}\\
&  \left[  \frac{\partial}{\partial t}+\mathbf{v}\cdot\frac{\partial}%
{\partial\mathbf{r}}-i\xi_{5}\right]  \rho_{g_{2}e_{2}}^{{}}-\gamma
_{\text{vcc}}R_{g_{2}e_{2}}F\nonumber\\
&  =-iV_{2}^{\ast}n_{0}F, \label{Eq. A11e}%
\end{align}
where $\xi_{i}$ ($i=1-5$) are given in Eq. (\ref{Eq. 3}). The expectation
value of the polarization density $\mathbf{P}_{e_{1}g_{2}}\left(
\mathbf{r},t\right)  $ in terms of the number density $R_{e_{1}g_{2}}\left(
\mathbf{r},t\right)  $ is $\mathbf{P}_{e_{1}g_{2}}\left(  \mathbf{r},t\right)
=\mu_{e_{1}g_{2}}^{\ast}R_{e_{1}g_{2}}\left(  \mathbf{r},t\right)  $, and Eq.
(\ref{Eq. A3}) becomes%

\end{subequations}
\begin{equation}
\left(  \frac{\partial}{\partial t}+c\frac{\partial}{\partial z}-i\frac
{c}{2q_{p}}\nabla_{\bot}^{2}\right)  V_{p}\left(  \mathbf{r},t\right)
=igR_{e_{1}g_{2}}\left(  \mathbf{r},t\right)  . \label{Eq. A12}%
\end{equation}

We now consider the case of stationary plane-wave pumps. For this case, it is
convenient to introduce the Fourier transform%

\begin{equation}
f\left(  \mathbf{r},t\right)  =\underset{-\infty}{\overset{+\infty}{\int}%
}\frac{d^{3}k}{2\pi}e^{i\mathbf{kr}}\underset{-\infty}{\overset{+\infty}{\int
}}\frac{d\omega}{2\pi}e^{-i\omega t}f\left(  \mathbf{k},\omega\right)  ,
\label{Eq. A13}%
\end{equation}
and write Eqs. (\ref{Eq. A11}) as%

\begin{subequations}
\label{Eq. A14}%
\begin{align}
&  \left[  \omega-\mathbf{k\cdot v}+\xi_{1}\right]  \rho_{g_{1}g_{2}}%
-i\gamma_{\text{vcc}}R_{g_{1}g_{2}}\left(  \mathbf{k},\omega\right)
F\nonumber\\
&  =\left(  V_{2}\rho_{g_{1}e_{2}}-V_{1}^{\ast}\rho_{e_{1}g_{2}}\right)
+ibA\Gamma\rho_{e_{1}e_{2}},\label{Eq. A14a}\\
&  \left[  \omega-\mathbf{k\cdot v}+\xi_{2}\right]  \rho_{e_{1}g_{2}}%
-i\gamma_{\text{vcc}}R_{e_{1}g_{2}}\left(  \mathbf{k},\omega\right)
F\nonumber\\
&  =-\left(  V_{p}n_{0}F+V_{1}\rho_{g_{1}g_{2}}\right)  ,\label{Eq. A14b}\\
&  \left[  \omega-\mathbf{k\cdot v}+\xi_{3}\right]  \rho_{e_{1}e_{2}}%
-i\gamma_{\text{vcc}}R_{e_{1}e_{2}}\left(  \mathbf{k},\omega\right)
F\nonumber\\
&  =\left(  V_{2}^{\ast}\rho_{e_{1}g_{2}}-V_{1}\rho_{g_{1}e_{2}}\right)
-V_{p}^{{}}\rho_{g_{2}e_{2}},\label{Eq. A14c}\\
&  \left[  \omega-\mathbf{k\cdot v}+\xi_{4}\right]  \rho_{g_{1}e_{2}}%
-i\gamma_{\text{vcc}}R_{g_{1}e_{2}}\left(  \mathbf{k},\omega\right)
F\nonumber\\
&  =V_{2}^{\ast}\rho_{g_{1}g_{2}},\label{Eq. A14d}\\
&  \left[  \omega-\mathbf{k\cdot v}+\xi_{5}\right]  \rho_{g_{2}e_{2}}%
-i\gamma_{\text{vcc}}R_{g_{2}e_{2}}\left(  \mathbf{k},\omega\right)
F\nonumber\\
&  =V_{2}^{\ast}n_{0}F, \label{Eq. A14e}%
\end{align}
and Eq. (\ref{Eq. A12}) as%

\end{subequations}
\begin{equation}
\left(  ik_{z}-i\frac{\omega}{c}+i\frac{k^{2}}{2q_{p}}\right)  V_{p}\left(
\mathbf{k},\omega\right)  =i\frac{g}{c}R_{e_{1}g_{2}}\left(  \mathbf{k}%
,\omega\right)  . \label{Eq. A15}%
\end{equation}
The linear susceptibility $\chi_{e_{1}g_{2}}\left(  \mathbf{k},\omega\right)
$ is defined by
\begin{equation}
R_{e_{1}g_{2}}\left(  \mathbf{k},\omega\right)  =\chi_{e_{1}g_{2}}\left(
\mathbf{k},\omega\right)  \frac{c}{g}V_{p}\left(  \mathbf{k},\omega\right)  .
\label{Eq. A16}%
\end{equation}

In order to find the probe absorption spectrum, we solve Eqs. (\ref{Eq. A14})
analytically, obtain an expression for $\rho_{ss^{\prime}}$, and formally
integrate it over velocity. This leads to an expression for $R_{ss^{\prime}}$
in terms of integrals over velocity, in the form of Eq. (\ref{Gs}), such as
$G_{1}=\int d^{3}v\frac{\xi_{2}\xi_{3}\xi_{4}F(\mathbf{v})}{\xi_{d}}$, which
can be evaluated numerically. In the general case, the resulting expression
for $R_{ss^{\prime}}$ is very complicated and is not reproduced here. In order
to explore the underlying physics, we developed an approximate expression for
the Fourier transform of the density-matrix element that refers to the probe
transition, namely, $R_{e_{1}g_{2}}$ [see Eq. (\ref{Eq. 2})].

One can verify that in the absence of the pumps ($V_{1}=V_{2}=0$), the
resulting one-photon complex spectrum simplifies to the well known result for
the strong collision regime, $K=iG/\left(  1-i\gamma_{\text{vcc}}G\right)  $,
where $G=\int d^{3}\mathbf{v}F /\left(  \omega-\mathbf{k\cdot v}+\xi
_{2}\right)  $ \cite{Sobelman1967SPU}.

\section{Diffusion in the presence of fields}

In order to obtain diffusion-like equations for the density-matrix elements
and the probe fields, we begin by integrating Eqs. (\ref{Eq. A11a}) and
(\ref{Eq. A11c}) over velocity and obtain%

\begin{subequations}
\label{Eq. B1}%
\begin{align}
&  \left[  \frac{\partial}{\partial\mathbf{r}}+i\delta\mathbf{q}_{1}\right]
\cdot\mathbf{J}_{g_{1}g_{2}}+\left[  \frac{\partial}{\partial t}-i\left(
\Delta_{p}-\Delta_{1}\right)  +\gamma\right]  R_{g_{1}g_{2}}\nonumber\\
&  =i\left(  V_{1}^{\ast}R_{e_{1}g_{2}}-V_{2}R_{g_{1}e_{2}}\right)  +bA\Gamma
R_{e_{1}e_{2}},\label{Eq. B1a}\\
&  \left[  \frac{\partial}{\partial\mathbf{r}}+i\delta\mathbf{q}_{2}\right]
\cdot\mathbf{J}_{e_{1}e_{2}}+\Biggl[\frac{\partial}{\partial t}-i\left(
\Delta_{p}-\Delta_{2}\right) \nonumber\\
&  +\Gamma+\gamma\Biggr]R_{e_{1}e_{2}}=i\left(  V_{1}R_{g_{1}e_{2}}%
-V_{2}^{\ast}R_{e_{1}g_{2}}+V_{p}R_{g_{2}e_{2}}\right)  , \label{Eq. B1b}%
\end{align}
where $\mathbf{J}_{ss^{\prime}}=\mathbf{J}_{ss^{\prime}}\left(  \mathbf{r}%
,t\right)  =\int d^{3}v\mathbf{v}\rho_{ss^{\prime}} $ is the envelope of the
current density. Expanding $\rho_{g_{1}g_{2}} $ and $\rho_{e_{1}e_{2}} $ in
Eqs. (\ref{Eq. A11a}) and (\ref{Eq. A11c}) as $\rho_{ss^{\prime}}
=R_{ss^{\prime}} F +1/\gamma_{\text{vcc}}\rho_{ss^{\prime}}^{(1)} ,$
multiplying Eqs. (\ref{Eq. A11a}) and (\ref{Eq. A11c}) by $\mathbf{v}$,
integrating the resulting equations over velocity using%

\end{subequations}
\begin{equation}
\int d^{3}v_{j}v_{i}\frac{\partial}{\partial x_{i}}R_{ss^{\prime}} F
=\delta_{ij}v_{\text{th}}\frac{\partial}{\partial x_{i}}R_{ss^{\prime}} ,
\label{Eq. B2}%
\end{equation}
defining the current density of the density matrix by%

\begin{equation}
\gamma_{\text{vcc}}\mathbf{J}_{ss^{\prime}} =\int d^{3}v_{j}\rho_{ss^{\prime}%
}^{\left(  1\right)  } , \label{Eq. B3}%
\end{equation}
and retaining the leading terms in $1/\gamma_{\text{vcc}}$, we obtain%

\begin{subequations}
\label{Eq. B4}%
\begin{align}
&  \mathbf{J}_{g_{1}g_{2}}+D\left[  \frac{\partial}{\partial\mathbf{r}%
}+i\delta\mathbf{q}_{1}\right]  R_{g_{1}g_{2}}\nonumber\\
&  =\frac{i}{\gamma_{\text{vcc}}}\left(  V_{1}^{\ast}\mathbf{J}_{e_{1}g_{2}%
}-V_{2}\mathbf{J}_{g_{1}g_{2}}\right)  -\frac{bA\Gamma}{\gamma_{\text{vcc}}%
}\mathbf{J}_{e_{1}e_{2}},\label{Eq. B4a}\\
&  \mathbf{J}_{e_{1}e_{2}}+D\left[  \frac{\partial}{\partial\mathbf{r}%
}+i\delta\mathbf{q}_{2}\right]  R_{e_{1}e_{2}}\nonumber\\
&  =\frac{i}{\gamma_{\text{vcc}}}\left(  V_{1}\mathbf{J}_{g_{1}e_{2}}%
-V_{2}^{\ast}\mathbf{J}_{e_{1}g_{2}}+\tilde{V}_{p}\mathbf{J}_{g_{2}e_{2}%
}\right)  , \label{Eq. B4b}%
\end{align}
where $D=v_{\text{th}}/\gamma_{\text{vcc}}$. Substituting $\mathbf{J}%
_{g_{1}g_{2}}$, $\mathbf{J}_{e_{1}e_{2}}$ from Eq. (\ref{Eq. B4}) into Eq.
(\ref{Eq. B1}), we get%

\end{subequations}
\begin{subequations}
\label{Eq. B5}%
\begin{align}
&  \left[  \frac{\partial}{\partial t}-i\left(  \Delta_{p}-\Delta_{1}\right)
+\gamma-D\left(  \frac{\partial}{\partial\mathbf{r}}+i\delta\mathbf{q}%
_{1}\right)  ^{2}\right]  R_{g_{1}g_{2}}\nonumber\\
&  =i\left(  V_{1}^{\ast}R_{e_{1}g_{2}}-V_{2}R_{g_{1}e_{2}}\right)  +bA\Gamma
R_{e_{1}e_{2}}-D\left(  \frac{\partial}{\partial\mathbf{r}}+i\delta
\mathbf{q}_{1}\right) \nonumber\\
&  \times\left[  \frac{i}{\gamma_{\text{vcc}}}\left(  V_{1}^{\ast}%
\mathbf{J}_{e_{1}g_{2}}-V_{2}\mathbf{J}_{g_{1}e_{2}}\right)  -\frac{bA\Gamma
}{\gamma_{\text{vcc}}}\mathbf{J}_{e_{1}e_{2}}\right]  ,\label{Eq. B5a}\\
&  \left[  \frac{\partial}{\partial t}-i\left(  \Delta_{p}-\Delta_{2}\right)
+\Gamma+\gamma-D\left(  \frac{\partial}{\partial\mathbf{r}}+i\delta
\mathbf{q}_{2}\right)  ^{2}\right]  R_{e_{1}e_{2}}\nonumber\\
&  =i\left(  V_{1}R_{g_{1}e_{2}}-V_{2}^{\ast}R_{e_{1}g_{2}}+V_{p}R_{g_{2}%
e_{2}}\right)  -D\left(  \frac{\partial}{\partial\mathbf{r}}+i\delta
\mathbf{q}_{2}\right) \nonumber\\
&  \times\left[  \frac{i}{\gamma_{\text{vcc}}}\left(  V_{1}\mathbf{J}%
_{g_{1}e_{2}}-V_{2}^{\ast}\mathbf{J}_{e_{1}g_{2}}+V_{p}\mathbf{J}_{g_{2}e_{2}%
}\right)  \right]  . \label{Eq. B5b}%
\end{align}
In order to calculate $R_{e_{1}g_{2}}$, $R_{g_{1}e_{2}}$, $R_{g_{2}e_{2}},$
and $\mathbf{J}_{e_{1}g_{2}}$, $\mathbf{J}_{g_{1}e_{2}}$, $\mathbf{J}%
_{g_{2}e_{2}}$, we assume in Eqs. (\ref{Eq. A11b}), (\ref{Eq. A11d}), and
(\ref{Eq. A11e}) that the envelopes change slowly enough such that $\left\vert
\partial/\partial t+\mathbf{v}\cdot\partial/\partial\mathbf{r}\right\vert $
$\ll\left\vert \xi_{2,4,5}\right\vert ,$ and get%

\end{subequations}
\begin{subequations}
\label{Eq. B6}%
\begin{align}
-i\xi_{2}\rho_{e_{1}g_{2}}=  &  \gamma_{\text{vcc}}R_{e_{1}g_{2}} F+i\left(
V_{p}n_{0}F+V_{1}\rho_{g_{1}g_{2}}\right)  ,\label{Eq. B6a}\\
-i\xi_{4}\rho_{g_{1}e_{2}}=  &  \gamma_{\text{vcc}}R_{g_{1}e_{2}}
F-iV_{2}^{\ast}\rho_{g_{1}g_{2}},\label{Eq. B6b}\\
-i\xi_{5}\rho_{g_{2}e_{2}}=  &  \gamma_{\text{vcc}}R_{g_{2}e_{2}}
F-iV_{2}^{\ast}n_{0}F. \label{Eq. B6c}%
\end{align}
Solving Eq. (\ref{Eq. B6}) formally for $\rho_{e_{1}g_{2}}$, $\rho_{g_{1}%
e_{2}}$, $\rho_{g_{2}e_{2}}$ and substituting only their leading parts,
\textit{i.e.} $\rho_{ss^{\prime}}=$ $R_{ss^{\prime}} F $, we find%

\end{subequations}
\begin{subequations}
\label{Eq. B7}%
\begin{align}
\rho_{e_{1}g_{2}}  &  =\left[  \gamma_{\text{vcc}}R_{e_{1}g_{2}} \right.
-V_{1}R_{g_{1}g_{2}}) - \left.  V_{p} n_{0}\right]  F/\xi_{2}, \label{Eq. B7a}%
\\
\rho_{g_{1}e_{2}}  &  =\left[  \gamma_{\text{vcc}}R_{g_{1}e_{2}} +V_{2}^{\ast
}R_{g_{1}g_{2}} \right]  F/\xi_{4},\label{Eq. B7b}\\
\rho_{g_{2}e_{2}}  &  =\left[  \gamma_{\text{vcc}}R_{g_{2}e_{2}} +V_{2}^{\ast
}n_{0}\right]  F/\xi_{5}. \label{Eq. B7c}%
\end{align}
Integrating Eqs. (\ref{Eq. B7}) over velocity we get%

\end{subequations}
\begin{subequations}
\label{Eq. B8}%
\begin{align}
&  R_{e_{1}g_{2}} =iK_{\text{1p}}\left[  V_{1}R_{g_{1}g_{2}} +V_{p}
n_{0}\right]  ,\label{Eq. B8a}\\
&  R_{g_{1}e_{2}} =-iK_{\text{3p}}^{{}}V_{2}^{\ast}R_{g_{1}g_{2}}
,\label{Eq. B8b}\\
&  R_{g_{2}e_{2}} =-iK_{\text{pump}}V_{2}^{\ast}n_{0}, \label{Eq. B8c}%
\end{align}
where $K_{\text{1p}}=iG_{\text{1p}}/\left(  1-G_{\text{1p}}\gamma_{\text{vcc}%
}\right)  $ is the one-photon absorption spectrum with $G_{\text{1p}}=\int F
/\xi_{2}d^{3}v\mathbf{\ }$, $K_{\text{3p}}=iG_{\text{3p}}/\left(
1-G_{\text{3p}}\gamma_{\text{vcc}}\right)  $ is the three-photon absorption
spectrum with $G_{\text{3p}}=\int F/\xi_{4}d^{3}v\mathbf{\ }$and
$K_{\text{pump}}=iG_{\text{pump}}/\left(  1-G_{\text{pump}}\gamma_{\text{vcc}%
}\right)  $ is the one-photon (pump) absorption spectrum with $G_{\text{pump}%
}=\int F/\xi_{5}d^{3}\mathbf{v}$. In the case of collinear pump and probe
beams $\delta\mathbf{q=}\delta\mathbf{q}_{1,2}=\mathbf{q}_{p}-\mathbf{q}%
_{1,2}=\delta q\widehat{\mathbf{z}}$, Eqs. (\ref{Eq. B5}) and (\ref{Eq. B8})
form a closed set when%

\end{subequations}
\begin{align*}
&  \left(  \frac{\partial}{\partial\mathbf{r}}+i\delta\mathbf{q}_{1,2}\right)
\cdot\frac{iV_{1,2}\left(  \mathbf{r}\right)  }{\gamma_{\text{vcc}}}%
\mathbf{J}_{e_{1}g_{2}},\\
&  \left(  \frac{\partial}{\partial\mathbf{r}}+i\delta\mathbf{q}_{1,2}\right)
\cdot\frac{iV_{1,2}\left(  \mathbf{r}\right)  }{\gamma_{\text{vcc}}}%
\mathbf{J}_{g_{1}e_{2}},\text{ }\\
&  \left(  \frac{\partial}{\partial\mathbf{r}}+i\delta\mathbf{q}_{2}\right)
\cdot\frac{iV_{p}\left(  \mathbf{r,}t\right)  }{\gamma_{\text{vcc}}}%
\mathbf{J}_{g_{2}e_{2}}%
\end{align*}
can be neglected in Eq. (\ref{Eq. B5}). These terms vanish completely in the
special case of pump and probe which are plane waves $\left(  \partial
/\partial\mathbf{r=0}\right)  $, and also collinear and degenerate $\left(
\delta\mathbf{q=0}\right)  $. They can also be neglected whenever $\left\vert
V_{1,2,p}\right\vert \ll\gamma_{\text{vcc}}$ as is the case in many realistic
situations. However, the term $\left(  \partial/\partial\mathbf{r}%
+i\delta\mathbf{q}_{1}\right)  \cdot bA\Gamma/\gamma_{\text{vcc}}%
\mathbf{J}_{e_{1}e_{2}}$ in Eq. (\ref{Eq. B5a}) cannot be neglected in the
case of collinear pump and probe beams since $bA\Gamma/\gamma_{\text{vcc}}$
does not go to zero.

Substituting Eq. (\ref{Eq. B4b}) into Eq. (\ref{Eq. B5a}), and Eq.
(\ref{Eq. B8}) into Eq. (\ref{Eq. B5}), we find:%

\begin{subequations}
\label{Eq. B9}%
\begin{align}
&  \left\{  \frac{\partial}{\partial t}-i\left(  \Delta_{p}-\Delta_{1}\right)
+\gamma+K_{\text{1p}}\left\vert V_{1}\right\vert ^{2}+K_{\text{3p}}\left\vert
V_{2}\right\vert ^{2}\right\}  R_{g_{1}g_{2}}\nonumber\\
&  =D\left(  \frac{\partial}{\partial\mathbf{r}}+i\delta\mathbf{q}_{1}\right)
^{2}R_{g_{1}g_{2}}+D\left(  \frac{\partial}{\partial\mathbf{r}}+i\delta
\mathbf{q}_{2}\right)  ^{2}R_{e_{1}e_{2}}\nonumber\\
&  +bA\Gamma R_{e_{1}e_{2}}-K_{\text{1p}}V_{1}^{\ast}V_{p}n_{0}%
,\label{Eq. B9a}\\
&  \left\{  \frac{\partial}{\partial t}-i\left(  \Delta_{p}-\Delta_{2}\right)
+\Gamma+\gamma\right\}  R_{e_{1}e_{2}}\nonumber\\
&  =D\left(  \frac{\partial}{\partial\mathbf{r}}+i\delta\mathbf{q}_{2}\right)
^{2}R_{e_{1}e_{2}}+V_{1}V_{2}^{\ast}\left(  K_{\text{1p}}+K_{\text{3p}%
}\right)  R_{g_{1}g_{2}}\nonumber\\
&  +V_{2}^{\ast}\left(  K_{\text{1p}}+K_{\text{pump}}\right)  V_{p}n_{0}.
\label{Eq. B9b}%
\end{align}
These are the final diffusion-like coupled equations for the ground- and
excited-state coherences.

In order to investigate the Ramsey narrowing of the EIA peak, we consider
finite probe and pump beams and restrict the discussion to collinear EIA. We
assume that the fields are stationary and overlap in their cross sections with
negligible variation along the $z-$direction, $V_{p}\left(  \mathbf{r}%
,t\right)  =V_{p}w\left(  \mathbf{r}_{\bot}\right)  $, $V_{1}\left(
\mathbf{r}\right)  =V_{1}w\left(  \mathbf{r}_{\bot}\right)  $, $V_{2}\left(
\mathbf{r},t\right)  =V_{2}w\left(  \mathbf{r}_{\bot}\right)  ,$ where
$w\left(  \mathbf{r}_{\bot}\right)  $ is the transverse profile of the fields.
We further take $\delta q=0$ and $\Delta_{1}=\Delta_{2}=0$ for brevity. In the
diffusion regime, we rewrite Eqs. (\ref{Eq. B8}) and (\ref{Eq. B9}) as%

\end{subequations}
\begin{subequations}
\label{Eq. B10}%
\begin{align}
&  \left[  i\Delta_{p}+\gamma+\left(  K_{\text{1p}}\left\vert V_{1}\right\vert
^{2}+K_{\text{3p}}\left\vert V_{2}\right\vert ^{2}\right)  w\left(
\mathbf{r}_{\bot}\right)  ^{2}\right]  R_{g_{1}g_{2}}=\nonumber\\
&  bA\Gamma\left(  1+\frac{D}{\gamma_{\text{vcc}}}\nabla_{\bot}^{2}\right)
R_{e_{1}e_{2}}-K_{\text{1p}}V_{1}^{\ast}V_{p}n_{0}w\left(  \mathbf{r}_{\bot
}\right)  ^{2},\label{Eq. B10a}\\
&  R_{e_{1}g_{2}}=iK_{\text{1p}}\left(  V_{1}R_{g_{1}g_{2}}+V_{p}n_{0}\right)
w\left(  \mathbf{r}_{\bot}\right)  ,\label{Eq. B10b}\\
&  \left(  i\Delta_{p}+\Gamma+\gamma-D\nabla_{\bot}^{2}\right)  R_{e_{1}e_{2}%
}\nonumber\\
&  =V_{1}\left(  K_{\text{1p}}+K_{\text{3p}}\right)  R_{g_{1}g_{2}}V_{2}%
^{\ast}w\left(  \mathbf{r}_{\bot}\right)  ^{2}\nonumber\\
&  +V_{p}\left(  K_{\text{1p}}+K_{\text{pump}}\right)  n_{0}V_{2}^{\ast
}w\left(  \mathbf{r}_{\bot}\right)  ^{2},\label{Eq. B10c}\\
&  R_{g_{1}e_{2}}=-iK_{\text{3p}}V_{2}^{\ast}R_{g_{1}g_{2}}w\left(
\mathbf{r}_{\bot}\right)  ,\label{Eq. B10d}\\
&  R_{g_{1}e_{2}}=-iK_{\text{pump}}V_{2}^{\ast}n_{0}w\left(  \mathbf{r}_{\bot
}\right)  . \label{Eq. B10e}%
\end{align}
We further consider a probe and pump beams with a uniform intensity and phase
within a sheet of thickness $2a$ in the $x-$direction (one-dimensional
stepwise beams):%

\end{subequations}
\[
w\left(  x,y\right)  =\left\{
\genfrac{}{}{0pt}{}{1\text{ for }\left\vert x\right\vert \leq a}{0\text{ for
}\left\vert x\right\vert >a}%
\right.  .
\]
The solution for $R_{g_{1}g_{2}}$, symmetric in $x$ and decaying as
$\left\vert x\right\vert \rightarrow\infty$, is given by%

\begin{subequations}
\label{Eq. B13}%
\begin{align}
&  R_{g_{1}g_{2}}\left(  \left\vert x\right\vert \leq a\right)  =\nonumber\\
&  C_{2}\cosh\left(  k_{1}x\right)  +C_{1}\cosh\left(  k_{2}x\right)
+\frac{bA\Gamma\beta_{2}+D\alpha_{2}^{2}\beta_{1}}{\left(  D\alpha_{1}%
\alpha_{2}\right)  ^{2}+bA\Gamma\beta_{3}},\label{Eq. B13a}\\
&  R_{e_{1}e_{2}}\left(  \left\vert x\right\vert \leq a\right)  =\frac
{C_{1}\left(  k_{2}^{2}-\alpha_{2}^{2}\right)  D\gamma_{\text{vcc}}}%
{bA\Gamma\left(  D\alpha_{2}^{2}+\gamma_{\text{vcc}}\right)  }\cosh\left(
k_{2}x\right) \nonumber\\
&  +\frac{C_{2}\left(  k_{1}^{2}-\alpha_{1}^{2}\right)  D\gamma_{\text{vcc}}%
}{bA\Gamma\left(  D\alpha_{2}^{2}+\gamma_{\text{vcc}}\right)  }\cosh\left(
k_{1}x\right)  +\frac{\beta_{1}\beta_{3}-\beta_{2}\alpha_{1}^{2}D}{\beta
_{3}bA\Gamma-\left(  D\alpha_{1}\alpha_{2}\right)  ^{2}},\\
&  R_{g_{1}g_{2}}\left(  \left\vert x\right\vert >a\right)  =\nonumber\\
&  \frac{C_{3}bA\Gamma\left(  D\alpha_{2}^{2}+\gamma_{\text{vcc}}\right)
}{\left(  \alpha_{3}^{2}-\alpha_{2}^{2}\right)  D\gamma_{\text{vcc}}%
}e^{-\alpha_{2}\left(  \left\vert x\right\vert -a\right)  }+C_{4}%
e^{-\alpha_{3}\left(  \left\vert x\right\vert -a\right)  },\\
&  R_{e_{1}e_{2}}\left(  \left\vert x\right\vert >a\right)  =C_{3}%
e^{-\alpha_{2}\left(  \left\vert x\right\vert -a\right)  },
\end{align}
where $\alpha_{1}^{2}=(-i\Delta_{p}+\gamma+K_{\text{1p}}\left\vert
V_{1}\right\vert ^{2}+K_{\text{3p}}\left\vert V_{2}\right\vert ^{2})/D,$
$\alpha_{2}^{2}=\alpha_{3}^{2}+\Gamma/D,$ $\alpha_{3}^{2}=\left(  -i\Delta
_{p}+\gamma\right)  /D,$ and $\beta_{1}=V_{1}^{\ast}V_{p}K_{\text{1p}}n_{0},$
$\beta_{2}=V_{1}V_{2}^{\ast}(K_{\text{1p}}+K_{\text{3p}}),\beta_{3}%
=V_{2}^{\ast}V_{p}(K_{\text{1p}}+K_{\text{pump}})n_{0}.$ The complex diffusion
wave-numbers are obtained from%

\end{subequations}
\begin{align*}
&  2D\gamma_{\text{vcc}}k_{1,2}^{2}=D\gamma_{\text{vcc}}\alpha_{+}^{2}%
+\beta_{3}bA\Gamma\\
&  \mp\left[  (D\gamma_{\text{vcc}})^{2}\alpha_{-}^{4}+\beta_{3}%
bA\Gamma\left(  4+2D\gamma_{\text{vcc}}\alpha_{+}^{2}+\beta_{3}bA\Gamma
\right)  \right]  ^{1/2},
\end{align*}
with $\alpha_{\pm}^{2}=\alpha_{2}^{2}\pm\alpha_{1}^{2}.$ The coefficients
$C_{i}$ ($i=1-4$) are obtained from the continuity conditions of
$R_{ss^{\prime}}$ and $\left(  \partial/\partial x\right)  R_{ss^{\prime}}$at
$\left\vert x\right\vert =a$. From Eq. (\ref{Eq. B10b}) one finds%

\begin{align}
&  R_{e_{1}g_{2}}\left(  \left\vert x\right\vert \leq a\right)  =iK_{2}%
\Biggl[V_{1}\left(  C_{2}\cosh\left(  k_{1}x\right)  \right. \nonumber\\
&  +C_{1}\cosh\left(  k_{2}x\right)  +\left.  \frac{bA\Gamma\beta_{2}%
+D\alpha_{2}^{2}\beta_{1}}{\left(  D\alpha_{1}\alpha_{2}\right)  ^{2}%
+bA\Gamma\beta_{3}}\right)  +V_{p}n_{0}\Biggr] , \label{B15}%
\end{align}
and the energy absorption at frequency $\omega_{p}$ is finally calculated from
$P\left(  \Delta\right)  =(\hbar\omega_{p}/a)\text{Im}\int_{-a}^{a}%
dxR_{e_{1}g_{2}}\left(  x\right)  .$ Two examples for the resulting spectrum
are given in Fig. \ref{Fig. 7}.


\end{document}